\documentclass[review,12pt]{elsarticle}

\usepackage[misc]{ifsym}
\usepackage{xcolor}

\usepackage[caption=false]{subfig}
\usepackage{float}

\usepackage{multirow}
\usepackage{rotating}
\usepackage{gensymb}

\usepackage{enumitem}

\usepackage{hyperref}

\usepackage[small]{caption}

\usepackage{siunitx}
\usepackage{pifont}

\usepackage{mathtools,etoolbox}
\DeclarePairedDelimiterX{\abs}[1]{\lvert}{\rvert}{\ifblank{#1}{{}\cdot{}}{#1}}

\usepackage{graphicx}
\usepackage{amssymb}
\usepackage{amsmath}

\usepackage{soul}
\usepackage{manyfoot}

\DeclareNewFootnote{A}
\DeclareNewFootnote{B}

\makeatletter
\def\Hline{
\noalign{\ifnum0=`}\fi\hrule \@height 1pt \futurelet
\reserved@a\@xhline}
\makeatother

\biboptions{sort&compress}

\journal{Computerized Medical Imaging and Graphics}

\begin{document}


\begin{titlepage}
	\clearpage\thispagestyle{empty}
		
\begin{center}
\begin{Large}Unified Focal loss: Generalising Dice and \\cross entropy-based losses to handle class imbalanced medical image segmentation\end{Large}
\end{center}

\noindent Michael Yeung\textsuperscript{1}, \textit{E-mail}: \texttt{mjyy2@cam.ac.uk}\\
\begin{scriptsize}
Department of Radiology, University of Cambridge, Cambridge CB2 0QQ, United Kingdom\\
School of Clinical Medicine, University of Cambridge, Cambridge CB2 0SP, United Kingdom
\end{scriptsize}

\noindent Evis Sala, \textit{E-mail}: \texttt{es220@medschl.cam.ac.uk}\\
\begin{scriptsize}
Department of Radiology, University of Cambridge, Cambridge CB2 0QQ, United Kingdom\\
Cancer Research UK Cambridge Centre, University of Cambridge, Cambridge CB2 0RE, United Kingdom
\end{scriptsize}

\noindent Carola-Bibiane Sch\"{o}nlieb, \textit{E-mail}: \texttt{cbs31@cam.ac.uk}\\
\begin{scriptsize}
Department of Applied Mathematics and Theoretical Physics, University of Cambridge, Cambridge CB3 0WA, United Kingdom
\end{scriptsize}

\noindent Leonardo Rundo\footnote[1]{Corresponding author. \textit{Phone}: +39 089 964355.\\  \textit{Address}: Department of Radiology, University of Cambridge, Box 218, Cambridge Biomedical Campus, Cambridge CB2 0QQ, United Kingdom.\\\textit{Present address:} Department of Information and Electrical Engineering and Applied Mathematics (DIEM), University of Salerno, Fisciano (SA) 84084, Italy}, \textit{E-mails}: \texttt{lr495@cam.ac.uk, lrundo@unisa.it}\\
\begin{scriptsize}
Department of Radiology, University of Cambridge, Cambridge CB2 0QQ, United Kingdom\\
Cancer Research UK Cambridge Centre, University of Cambridge, Cambridge CB2 0RE, United Kingdom\\
Department of Information and Electrical Engineering and Applied Mathematics (DIEM),
University of Salerno, Fisciano (SA) 84084, Italy
\end{scriptsize}

\newpage
\end{titlepage}


\begin{frontmatter}


\title{Unified Focal loss: Generalising Dice and \\cross entropy-based losses to handle class imbalanced medical image segmentation}




\author[CamMedSchool,CamRadiol]{Michael Yeung\textsuperscript{*,}}\ead{mjyy2@cam.ac.uk}

\author[CamRadiol,CRUKCC]{Evis Sala}\ead{es220@medschl.cam.ac.uk}

\author[DAMTP]{Carola-Bibiane Sch\"{o}nlieb}\ead{cbs31@cam.ac.uk}

\author[CamRadiol,CRUKCC,UniSA]{Leonardo Rundo\corref{corr}}\ead{lr495@cam.ac.uk, lrundo@unisa.it}

\cortext[corr]{Corresponding author}

\address[CamRadiol]{Department of Radiology, University of Cambridge,\\Cambridge CB2 0QQ, United Kingdom}
\address[CamMedSchool]{School of Clinical Medicine, University of Cambridge,\\Cambridge CB2 0SP, United Kingdom}
\address[CRUKCC]{Cancer Research UK Cambridge Centre, University of Cambridge,\\Cambridge CB2 0RE, United Kingdom}
\address[DAMTP]{Department of Applied Mathematics and Theoretical Physics, University of Cambridge, Cambridge CB3 0WA, United Kingdom}
\address[UniSA]{Department of Information and Electrical Engineering and Applied Mathematics (DIEM), University of Salerno, Fisciano (SA) 84084, Italy}

\begin{abstract}

Automatic segmentation methods are an important advancement in medical image analysis. Machine learning techniques, and deep neural networks in particular, are the state-of-the-art for most medical image segmentation tasks. Issues with class imbalance pose a significant challenge in medical datasets, with lesions often occupying a considerably smaller volume relative to the background. Loss functions used in the training of deep learning algorithms differ in their robustness to class imbalance, with direct consequences for model convergence. The most commonly used loss functions for segmentation are based on either the cross entropy loss, Dice loss or a combination of the two. We propose the Unified Focal loss, a new hierarchical framework that generalises Dice and cross entropy-based losses for handling class imbalance. We evaluate our proposed loss function on five publicly available, class imbalanced  medical imaging datasets: CVC-ClinicDB, Digital Retinal Images for Vessel Extraction (DRIVE), Breast Ultrasound 2017 (BUS2017), Brain Tumour Segmentation 2020 (BraTS20) and Kidney Tumour Segmentation 2019 (KiTS19). We compare our loss function performance against six Dice or cross entropy-based loss functions, across 2D binary, 3D binary and 3D multiclass segmentation tasks, demonstrating that our proposed loss function is robust to class imbalance and consistently outperforms the other loss functions. Source code is available at: \url{https://github.com/mlyg/unified-focal-loss}.\end{abstract}



\end{frontmatter}


\section{Introduction}
\label{sec:Intro}

Image segmentation involves partitioning an image into meaningful regions, based on the regional pixel characteristics, from which objects of interest are identified \citep{pal1993review}. This is a fundamental task in computer vision and has been applied widely in face recognition, autonomous driving, as well as medical image processing. In particular, automatic segmentation methods are an important advancement in medical image analysis, capable of demarcating structures across a range of imaging modalities including ultrasound (US), computed tomography (CT) and magnetic resonance imaging (MRI).

Classical approaches for image segmentation include direct region detection methods (such as the split-and-merge and region growing algorithms \citep{rundoMBEC2016}), graph-based methods \citep{chen2018survey}, active contour and level set models \citep{khadidos2017weighted}. Later approaches have focused on applying and adapting traditional machine learning techniques \citep{rundo2019FI}, such as support vector machines (SVMs) \citep{wang2012}, unsupervised clustering \citep{ren2019study} and atlas-based segmentation \citep{wachinger2014atlas}. In recent years, however, significant progress has been achieved using deep learning \citep{ker2018,rueckert2019model,castiglioni2021}.

The most well-known architecture in image segmentation, the U-Net \citep{ronneberger2015u}, is a modification of the convolutional neural network (CNN) architecture into an encoder-decoder network, similar to SegNet \citep{badrinarayanan2017segnet}, which enables end-to-end feature extraction and pixel classification. Since its inception, many variants based on the U-Net architecture have been proposed \citep{liu2020survey,Rundo2019USEnet}---including the 3D U-Net \citep{cciccek20163d}, Attention U-Net \citep{schlemper2019} and V-Net \citep{milletari2016v}---as well as integrated into conditional Generative Adversarial Networks \citep{kessler2020,armanious2020medgan}. 

To train deep neural networks, backpropagation updates model parameters in accordance with the optimisation goal defined by the loss function. The cross entropy loss is typically the most widely used loss function in classification problems \citep{liu2020prostate} and is applied in the U-Net \citep{ronneberger2015u}, 3D U-Net \citep{cciccek20163d} and SegNet \citep{badrinarayanan2017segnet}. In contrast, the Attention U-Net \citep{schlemper2019} and V-Net \citep{milletari2016v} leverage the Dice loss, which is based on the most commonly used metric for evaluating segmentation performance, and therefore represents a form of direct loss minimisation. Broadly, loss functions used in image segmentation may be classified into distribution-based losses (such as the cross entropy loss), region-based losses (such as Dice loss), boundary-based losses (such as the boundary loss) \citep{kervadec2019boundary}, and more recently compound losses. Compound losses combine multiple, independent loss functions, such as the Combo loss, which is the sum of the Dice and cross entropy loss \citep{taghanaki2019combo}. 

A dominant issue in medical image segmentation is handling class imbalance, which refers to an unequal distribution of foreground and background elements. For example, automatic organ segmentation often involves organ sizes that are an order of magnitude smaller than the scan itself, resulting in a skewed distribution favouring background elements \citep{roth2015deeporgan}. This issue is even more prevalent in oncology, where tumour sizes are themselves often significantly smaller than the associated organ of origin. 

\cite{taghanaki2019combo} distinguish between input and output imbalance, the former as aforementioned, and the latter referring to classification biases arising during inference. These include false positives and false negatives, which respectively describe background pixels incorrectly classified as foreground objects, and foreground objects incorrectly classified as background. Both are particularly important in the context of medical image segmentation; in the case of image-guided interventions, false positives may result in a larger radiation field or excessive surgical margins, and conversely false negatives may lead to inadequate radiation delivery or incomplete surgical resection. Therefore, it is important to design a loss function that can be optimised to handle both input and output imbalances. 

Despite its significance, careful selection of the loss function is not widespread practice, and often suboptimal loss functions are chosen with performance repercussions. To inform loss function choice, it is important to perform large-scale loss function comparisons. Seven loss functions were compared on the CVC-EndoSceneStill (gastrointestinal polyp segmentation) dataset, with the best performance seen with region-based losses and conversely the worst performance with the cross entropy loss \citep{sanchez2020eigenloss}. Similarly, a comparison of fifteen loss functions using the NBFS Skull-stripping dataset \citep{eskildsen2012beast} (brain CT segmentation), which also introduces the log-cosh Dice loss, concluded that Focal Tversky loss and Tversky loss, both region-based losses, are generally optimal \citep{9277638}. This is further supported by the most comprehensive loss function comparison to the date, with twenty loss functions compared across four datasets (liver, liver tumour, pancreas and multi-organ segmentation), which observed the best performance with compound-based losses, where the most consistent performance was observed with the DiceTopK and DiceFocal loss \citep{ma2021loss}. It is apparent from these studies that region-based or compound losses are associated with consistently better performance than distribution-based losses. Less clear, however, is which of the region-based or compound losses to choose, with no agreement among the aforementioned. One major confounding factor is the degree of class imbalance in the datasets, with low class imbalance seen in the NBFS Skull-stripping dataset, moderate class imbalance in the CVC-EndoSceneStill dataset, and a combination of both low and high class imbalanced datasets present in \citep{ma2021loss}. 

Among medical imaging datasets, those involving tumour segmentation are associated with high degrees of class imbalance. Manual tumour delineation is both time-consuming and operator-dependent. Automatic methods of tumour delineation aim to address these issues, and public datasets, such as the Breast Ultrasound 2017 (BUS2017) dataset for breast tumours \citep{yap2017automated}, Kidney Tumour Segmentation 19 (KiTS19) dataset for kidney tumours \citep{heller2019kits19} and Brain Tumour Segmentation 2020 (BraTS20) for brain tumours \citep{menze2014multimodal}, have accelerated progress towards this goal. In fact, there has been recent developments for translating the BraTS20 dataset into clinical and scientific practice \citep{kofler2020brats}.

Current state-of-the-art models for the BUS2017 dataset incorporate attention gates, which may provide benefits in class imbalanced situations by using contextual information from the gating signal to refine skip connections, highlighting the regions of interest \citep{8759329}. In addition to attention gates, the RDAU-NET combines residual units and dilated convolutions to enhance information transfer and increase the receptive field, respectively, and was trained using the Dice loss \citep{zhuang2019rdau}. The multi-input Attention U-Net combines attention gates with deep supervision, and introduces the Focal Tversky loss, a region-based loss function designed to handle class imbalance \citep{8759329}.

For the BraTS20 dataset, a popular approach is to use a multi-scale architecture where different receptive field sizes allow for the independent processing of both local and global contextual information \citep{kamnitsas2017efficient,havaei2017brain}. \cite{kamnitsas2017efficient} used a two-phase training process involving initial upsampling of under-represented classes, followed by a second-stage where the output layer is retrained on a more representative sample. Similarly, \cite{havaei2017brain} used a sampling rule to impose equal probability of foreground or background pixels at the centre of a patch, and used the cross entropy loss for optimisation.

For the KiTS19 dataset, the current state-of-the-art is the ``no-new-Net'' (nnU-Net) \citep{isensee2020NatMeth,isensee2018nnu}, an automatically configurable deep learning-based segmentation method involving the ensemble of 2D, 3D and cascaded 3D U-Nets. This framework was optimised using the Dice and cross entropy loss. Recently, an ensemble-based method obtained comparable results to nnU-Net, and involved initial independent processing of kidney organ and kidney tumour segmentation by 2D U-Nets trained using the Dice loss, followed by suppression of false positive predictions of the kidney tumour segmentation using the network trained for kidney organ segmentation \citep{fatemeh2020ensemble}. When the dataset size is small, results from an active learning-based method using CNN-corrected labeling, also trained using the Dice loss, showed a higher segmentation accuracy over nnU-Net \citep{kim2020active}.

It is apparent that for all three datasets, class imbalance is largely handled by altering either the training or input data sampling process, and rarely with adapting the loss function. However, popular methods---such as upsampling the underrepresented class---are inherently associated with an increase in false positive predictions, and more complicated, often multi-stage training processes require more computational resources.

State-of-the-art solutions typically use unmodified versions of either the Dice loss, cross entropy loss or a combination of the two, and even when using available loss functions for handling class imbalance, such as the Focal Tversky loss, consistently improved performance has not been observed \citep{ma2021loss}. Deciding which loss function to use is difficult because there is not only a significant number of loss functions available to choose from, but it is also unclear how each loss function relates to one another. Understanding the relationship between loss functions is the key for providing heuristics to inform loss function choice in class imbalanced situations.

In this paper, we propose the following contributions:
\begin{enumerate}[label=(\alph*)]
\item	We summarise and extend the knowledge provided by previous studies that compare loss functions to address the context of class imbalance, by using five class imbalanced datasets with varying degrees of class imbalance, including 2D binary, 3D binary and 3D multi-class segmentation, across multiple imaging modalities. 
\item	We define a hierarchical classification of Dice and cross entropy-based loss functions, and use this to derive the Unified Focal loss, that generalises Dice-based and cross entropy-based loss functions for handling class imbalanced datasets.
\item	Our proposed loss function consistently improves segmentation quality over six other related loss functions, is associated with a better recall-precision balance, and is robust to class imbalance.
\end{enumerate}

The manuscript is organised as follows.
Section~\ref{sec:Background} provides a summary of the loss functions used, including the proposed Unified Focal loss.
Section~\ref{sec:MatMeth} describes the chosen medical imaging datasets and defines the segmentation evaluation metrics used.
Section~\ref{sec:Results} presents and discusses the experimental results.
Finally, Section~\ref{sec:Discussion} provides conclusive remarks and future directions.

\section{Background}
\label{sec:Background}

\begin{figure}[ht!]
    \centering
    \includegraphics[width=\textwidth]{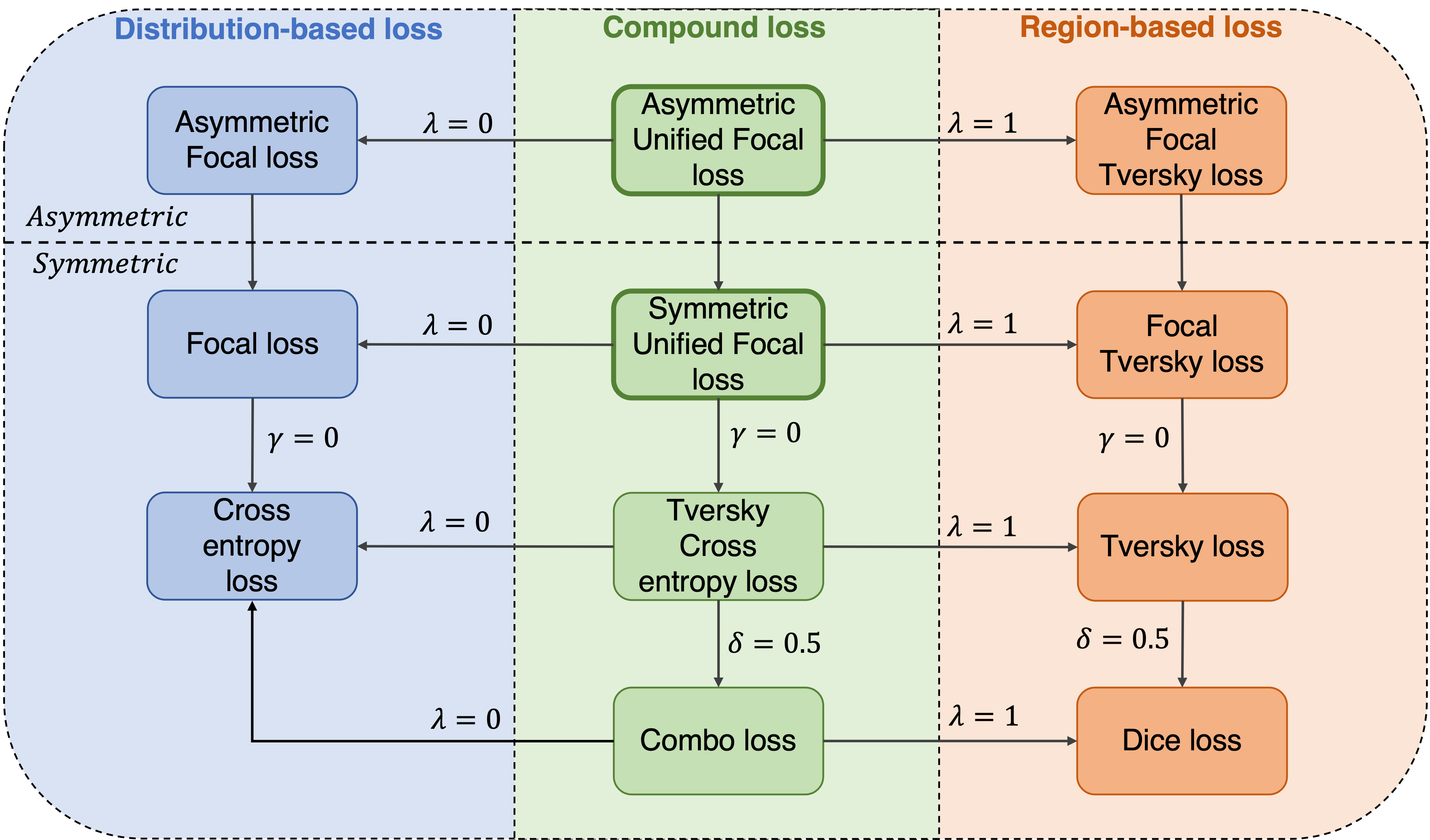}
    \caption{Our proposed framework unifying various distribution-based, region-based and compound loss functions. The arrows and associated hyperparameter values indicate the required hyperparameter value to set for the preceding loss function in order to recover the resulting loss function.}
    \label{fig:figure_1}
\end{figure}

The loss function defines the optimisation problem, and directly affects model convergence during training. This paper focuses on semantic segmentation, a sub-field of image segmentation where pixel-level classification is performed directly, in contrast to instance segmentation where an additional object detection stage is required. We describe seven loss functions that belong to either distribution-based, region-based or compound losses based of a combination of the two. A graphical overview of loss functions in these categories, and how all are derivable from the Unified Focal loss, is provided in Fig.~\ref{fig:figure_1}. First, the distribution-based functions are introduced, followed by region-based loss functions, and finally concluding with compound loss functions. 

\subsection{Cross entropy loss}

The cross entropy loss is one of the most widely used loss functions in deep learning. With origins in information theory, cross entropy measures the difference between two probability distributions for a given random variable or set of events. As a loss function, it is superficially equivalent to the negative log likelihood loss and, for binary classification, the binary cross entropy loss ($\mathcal{L}_\text{BCE}$) is defined as the following:

\begin{equation}
\mathcal{L}_\text{BCE}(\mathbf{y}, \hat{\mathbf{y}})=-(\mathbf{y} \log (\hat{\mathbf{y}})+(\mathbf{1}-\mathbf{y}) \log (\mathbf{1}-\hat{\mathbf{y}})).
\label{eq:BCE}
\end{equation}

Here, $\mathbf{y},\hat{\mathbf{y}}\in\{0,1\}^N$, where $\hat{\mathbf{y}}$ refers to the predicted value and $\mathbf{y}$ refers to the ground truth label. This can be extended to multi-class problems, and the categorical cross entropy loss ($\mathcal{L}_\text{CCE}$) is computed as:

\begin{equation}
\mathcal{L}_\text{CCE}(y, p)=-\frac{1}{N} \sum_{i=1}^{N} \sum_{c=1}^{C} y_{i, c} \cdot \log \left(p_{i, c}\right),
\label{eq:CCE}
\end{equation}
where $y_{i, c}$ uses a one-hot encoding scheme of ground truth labels, $p_{i, c}$ is a matrix of predicted values for each class, and where indices $c$ and $i$ iterate over all classes and pixels, respectively. Cross entropy loss is based on minimising pixel-wise error, where in class imbalanced situations, leads to over-representation of larger objects in the loss, resulting in poorer quality segmentation of smaller objects.

\subsection{Focal loss}

The Focal loss is a variant of the binary cross entropy loss that addresses the issue of class imbalance with the standard cross entropy loss by down-weighting the contribution of easy examples enabling learning of harder examples \citep{Lin_2017_ICCV}. To derive the Focal loss function, we first simplify the loss in Eq.~(\ref{eq:BCE}) as:

\begin{equation}
\text{CE}(p,y)=\begin{cases}
-\log (p), & \text { if } y=1 \\
-\log (1-p), & \text { if } y=0
\end{cases}.
\label{eq:CE}
\end{equation}

Next, we define the probability of predicting the ground truth class, $p_{t}$, as:

\begin{equation}
p_{t}=\begin{cases}
p, & \text { if } y=1 \\
1-p, & \text { if } y=0
\end{cases}.
\label{eq:predProb}
\end{equation}

The binary cross entropy loss ($\mathcal{L}_\text{BCE}$) can therefore be rewritten as:

\begin{equation}
\mathcal{L}_{\text{BCE}(p, y)} = \text{CE}\left(p_{t}\right)=-\log \left(p_{t}\right).
\label{eq:CE_predProb}
\end{equation}

The Focal loss ($\mathcal{L}_\text{F}$) adds a modulating factor to the binary cross entropy loss:

\begin{equation}
\mathcal{L}_{F\left(p_{t}\right)}=\alpha\left(1-p_{t}\right)^{\gamma} \cdot \mathcal{L}_{\text{BCE}(p, y)},
\label{eq:F_predProb}
\end{equation}

The Focal loss is parameterised by $\alpha$ and $\gamma$, which control the class weights and degree of down-weighting of easy-to-classify pixels, respectively (Fig.~\ref{fig:figure_2}). When $\gamma=0$, the Focal loss simplifies to the binary cross entropy loss. 

For multi-class segmentation, we define the categorical Focal loss ($\mathcal{L}_\text{CF}$):

\begin{equation}
\mathcal{L}_\text{CF}=\boldsymbol{\alpha}\left(1-\left(p_{t, c}\right)\right)^{\gamma} \cdot \mathcal{L}_\text{CCE},
\label{eq:CF}
\end{equation}
where $\boldsymbol{\alpha}$ is now a vector of class weights, $p_{t,c}$ is a matrix of ground truth probabilities for each class, and $\mathcal{L}_\text{CCE}$ is the categorical cross entropy loss as defined in Eq.~(\ref{eq:CCE}). 

\subsection{Dice loss}

The Sørensen–Dice index, known as the Dice similarity coefficient (DSC) when applied to Boolean data, is the most commonly used metric for evaluating segmentation accuracy. We can define DSC in terms of the per voxel classification of true positives (TP), false positives (FP) and false negatives (FN):

\begin{equation}
\text{DSC}=\frac{2 \text{TP}}{2 \text{TP} + \text{FP}+ \text{FN}}.
\label{eq:DSC}
\end{equation}

The Dice loss ($\mathcal{L}_\text{DSC}$), can therefore be defined as:

\begin{equation}
\mathcal{L}_{\text{DSC}}=1-\text{DSC}.
\label{eq:L_DSC}
\end{equation}

Other variants of the Dice loss include the Generalised Dice loss \citep{1717643,sudre2017generalised} where the class weights are corrected by the inverse of their volume, and the Generalised Wasserstein Dice loss \citep{fidon2017generalised}, which combines the Wasserstein metric with the Dice loss and is adapted for dealing with hierarchical data, such as the BraTS20 dataset \citep{menze2014multimodal}. 

Even in its most simple formulation, the Dice loss is somewhat adapted to handle class imbalance. However, the Dice loss gradient is inherently unstable, most evident with highly class imbalanced data where gradient calculations involve small denominators \citep{wong20183d, bertels2019optimization}.

\subsection{Tversky loss}

The Tversky index \citep{salehi2017tversky} is closely related to the DSC, but enables optimisation for output imbalance by assigning weights $\alpha$ and $\beta$ to false positives and false negatives respectively:

\begin{equation}
\operatorname{TI} =
\frac{\sum_{i=1}^{N} p_{0 i} g_{0 i}}{\sum_{i=1}^{N} p_{0 i} g_{0 i}+\alpha\sum_{i=1}^{N} p_{0 i} g_{1 i}+\beta \sum_{i=1}^{N} p_{1 i} g_{0 i}},
\label{eq:Tversky_index}
\end{equation}
where $p_{0 i}$ is the probability of pixel $i$ belonging to the foreground class and $p_{1 i}$ is the probability of pixel belonging to background class. $g_{0 i}$ is $1$ for foreground and 0 for background and conversely $g\textsubscript{1i}$ takes values of $1$ for background and 0 for foreground. 

Using the Tversky index, we define the Tversky loss ($\mathcal{L}_\text{T}$) for $C$ classes as:

\begin{equation}
\mathcal{L}_{T}=\sum_{c=1}^{C}(1-\text{TI})
\label{eq:Tversky}
\end{equation}

When the Dice loss function is applied to class imbalanced problems, the resulting segmentation often exhibits high precision but low recall scores \citep{salehi2017tversky}. Assigning a greater weight to false negatives improves recall and results in a better balance of precision and recall. Therefore, $\beta$ is often set higher than $\alpha$, most commonly $\beta = 0.7$ and $\alpha = 0.3$.

The asymmetric similarity loss is derived from the Tversky loss, but uses the $F_{\beta}$ score and substitutes $\alpha$ for $\frac{1}{1+\beta^2}$ and $\beta$ for $\frac{\beta^2}{1+\beta^2}$, adding the constraint that $\alpha$ and $\beta$ must sum to $1$ \citep{hashemi2018asymmetric}. In practice, $\alpha$ and $\beta$ values for the Tversky loss are chosen such that they sum to $1$, making both loss functions functionally equivalent. 

\subsection{Focal Tversky loss}

Inspired by the Focal loss adaptation of the cross entropy loss, the Focal Tversky loss \citep{8759329} adapts the Tversky loss by applying a focal parameter. 

Using the definition of TI from Eq.~(\ref{eq:Tversky_index}),the Focal Tversky loss is defined ($\mathcal{L}_\text{FT}$) as:

\begin{equation}
\mathcal{L}_\text{FT}=\sum_{c=1}^{C}(1-\text{TI})^{\frac{1}{\gamma}},
\label{eq:FT}
\end{equation}
where $\gamma < 1$ increases the degree of focusing on harder examples. The Focal Tversky loss simplifies to the Tversky loss when $\gamma = 1$. However, contrary to the Focal loss, the optimal value reported was $\gamma = 4/3$, which enhances rather than suppresses the loss of easy examples. Indeed, near the end of training where the majority of the examples are more confidently classified and the Tversky index approaches $1$, enhancing the loss in this region maintains a higher loss which may prevent premature convergence to a suboptimal solution. 

\subsection{Combo loss}

The Combo loss \citep{taghanaki2019combo} belongs to the class of compound losses, where multiple loss functions are minimised in unison. 
The Combo loss ($\mathcal{L}_\text{combo}$) is defined as a weighted sum of the DSC in Eq.~(\ref{eq:DSC}) and a modified form of the cross entropy loss ($\mathcal{L}_\text{mCE}$):

\begin{equation}
\mathcal{L}_\text{combo}=\alpha\left(\mathcal{L}_\text{mCE}\right)-(1-\alpha) \cdot {\text{DSC}},
\label{eq:combo}
\end{equation}
where: 
\begin{equation}
\mathcal{L}_\text{mCE}=-\frac{1}{N} \sum_{i=1}^{N} \beta\left(y_{i}\ln \left(p_{i}\right)\right)+(1-\beta)\left[\left(1-y_{i}\right) \ln \left(1-p_{i}\right)\right]
\label{eq:MCE}
\end{equation}
and $\alpha \in [0,1]$ controls the relative contribution of the Dice and cross entropy terms to the loss, and $\beta$ controls the relative weights assigned to false positives and negatives. A value of $\beta > 0.5$ penalises false negative predictions more than false positives.

Confusingly, the term ``Dice and cross entropy loss'' has been used to refer to both the sum of cross entropy loss and DSC \citep{taghanaki2019combo,isensee2018nnu}, as well as the sum of the cross entropy loss and Dice loss, such as in the DiceFocal loss and Dice and weighted cross entropy loss \citep{zhu2019anatomynet,chen2019robust}. Here, we decide to use the former definition, which is consistent with both Combo loss and the loss function used in the state-of-the-art for the KiTS19 dataset \citep{isensee2018nnu}.

\subsection{Hybrid Focal loss}
\label{sec:HybridFocalLoss}
The Combo loss \citep{taghanaki2019combo} and DiceFocal loss \citep{zhu2019anatomynet} are two compound loss functions that inherit benefits from both Dice and cross entropy-based loss functions. However, neither exploits the full benefits in the context of class imbalance. Both the Combo loss and the DiceFocal loss, with a tunable $\beta$ and $\alpha$ parameter respectively in the cross entropy component losses, are partially robust to output imbalance. However, both lack an equivalent for the Dice component loss, where positive and negative examples remain equally weighted. Similarly, the Dice component of both losses are not adapted to handle input imbalance, although the DiceFocal loss is better adapted with its focal parameter in the Focal loss component. 

To overcome this, we previously proposed the Hybrid Focal loss function, which incorporates tunable parameters to handle output imbalance, as well as focal parameters to handle input imbalance, for both the Dice and cross entropy-based component losses \citep{yeung2021FocusUnet}. By replacing the Dice loss with the Focal Tversky loss, and the cross entropy loss with the Focal loss, the Hybrid Focal loss ($\mathcal{L}_\text{HF}$) is defined as:

\begin{equation}
\mathcal{L}_\text{HF}=\lambda \mathcal{L}_\text{F}+(1-\lambda) \mathcal{L}_\text{FT},
\label{eq:BMF}
\end{equation}
where $\lambda \in [0,1]$ and determines the relative weighting of the two component loss functions.

\subsection{Unified Focal loss}
\label{sec:UnifiedFocalLoss}

The Hybrid Focal loss adapts both the Dice and cross entropy based losses to handle class imbalance. However, there are two main issues associated with using the Hybrid Focal loss in practice. Firstly, there are six hyperparameters to tune: $\alpha$ and $\gamma$ from the Focal loss, $\alpha$ / $\beta$ and $\gamma$ from the Focal Tversky loss, and $\lambda$ to control the relative weighting of the two component losses. While this allows a greater degree of flexibility, this comes at the cost of a significantly larger hyperparameter search space. The second issue is common to all focal loss functions, where the enhancing or suppressing effect introduced by the focal parameter is applied to all classes, which may affect the convergence towards the end of training.

The Unified Focal loss addresses both issues, by grouping functionally equivalent hyperparameters together and exploiting asymmetry to focus the suppressive and enhancing effects of the focal parameters in the modified Focal loss and Focal Tversky loss components, respectively.

Firstly, we replace $\alpha$ in the Focal loss and $\alpha$ and $\beta$ in the Tversky Index with a common $\delta$ parameter to control output imbalance, and reformulate $\gamma$ to enable simultaneous Focal loss suppression and Focal Tversky loss enhancement, naming these the modified Focal loss ($\mathcal{L}_\text{mF}$) and modified Focal Tversky loss ($\mathcal{L}_\text{mFT}$), respectively:

\begin{equation}
\mathcal{L}_{mF\left(p_{t}\right)}=\delta\left(1-p_{t}\right)^{1-\gamma} \cdot \mathcal{L}_{\text{BCE}(p, y)},
\label{eq:modified_Focal_loss}
\end{equation}

\begin{equation}
\mathcal{L}_\text{mFT}=\sum_{c=1}^{C}(1-\text{mTI})^{\gamma},
\label{eq:mFT}
\end{equation}

where,

\begin{equation}
\operatorname{mTI} =
\frac{\sum_{i=1}^{N} p_{0 i} g_{0 i}}{\sum_{i=1}^{N} p_{0 i} g_{0 i}+\delta\sum_{i=1}^{N} p_{0 i} g_{1 i}+(1-\delta) \sum_{i=1}^{N} p_{1 i} g_{0 i}}.
\label{eq:modified_Tversky_index}
\end{equation}

The symmetric variant of the Unified Focal loss ($\mathcal{L}_\text{sUF}$) is therefore defined as:

\begin{equation}
\mathcal{L}_\text{sUF}=\lambda \mathcal{L}_\text{mF}+(1-\lambda) \mathcal{L}_\text{mFT},
\label{eq:sUF}
\end{equation}

where $\lambda \in [0,1]$ and determines the relative weighting of the two losses. By grouping functionally equivalent hyperparameters, the six hyperparameters associated with the Hybrid Focal loss are reduced to three, with $\delta$ controlling the relative weighting of positive and negative examples, $\gamma$ controlling both suppression of the background class and enhancement of the rare class, and finally $\lambda$ determining the weights of the two component losses.

Although the Focal loss achieves suppression of the background class, the focal parameter is applied to all classes and therefore the loss contributed by the rare class is also suppressed. Asymmetry enables selective enhancement or suppression using the focal parameter by assigning different losses to each class, and this overcomes the both the harmful suppression of the rare class and enhancement of the background class. The modified asymmetric Focal loss ($\mathcal{L}_\text{maF}$) removes the focal parameter for the component of the loss relating to the rare class $r$, while retaining suppression of the background elements \citep{li2019overfitting}:

\begin{equation}
\mathcal{L}_\text{maF}=- \frac{\delta}{N} y_{i: r} \log \left(p_{t, r}\right)-\frac{1-\delta}{N} \sum_{c \neq r}\left(1-p_{t, c}\right)^{\gamma} \log \left(p_{t, r}\right),
\end{equation}

In contrast, for the modified Focal Tversky loss, we remove the focal parameter for the component of the loss relating to the background, retaining enhancement of the rare class $r$, and define the modified asymmetric Focal Tversky loss ($\mathcal{L}_\text{maFT}$) as:

\begin{equation}
\mathcal{L}_{\mathrm{maFT}}=\sum_{\mathrm{c} \neq \mathrm{r}}(1-\mathrm{mTI})+\sum_{\mathrm{c}=\mathrm{r}}(1-\mathrm{mTI})^{1-\gamma}
\end{equation}

The asymmetric variant of the Unified Focal loss ($\mathcal{L}_\text{aUF}$), is therefore defined as:

\begin{equation}
\mathcal{L}_\text{aUF}=\lambda \mathcal{L}_\text{maF}+(1-\lambda) \mathcal{L}_\text{maFT},
\label{eq:aUF}
\end{equation}

\begin{figure}[ht!]
    \centering
    \includegraphics[width=0.7\textwidth]{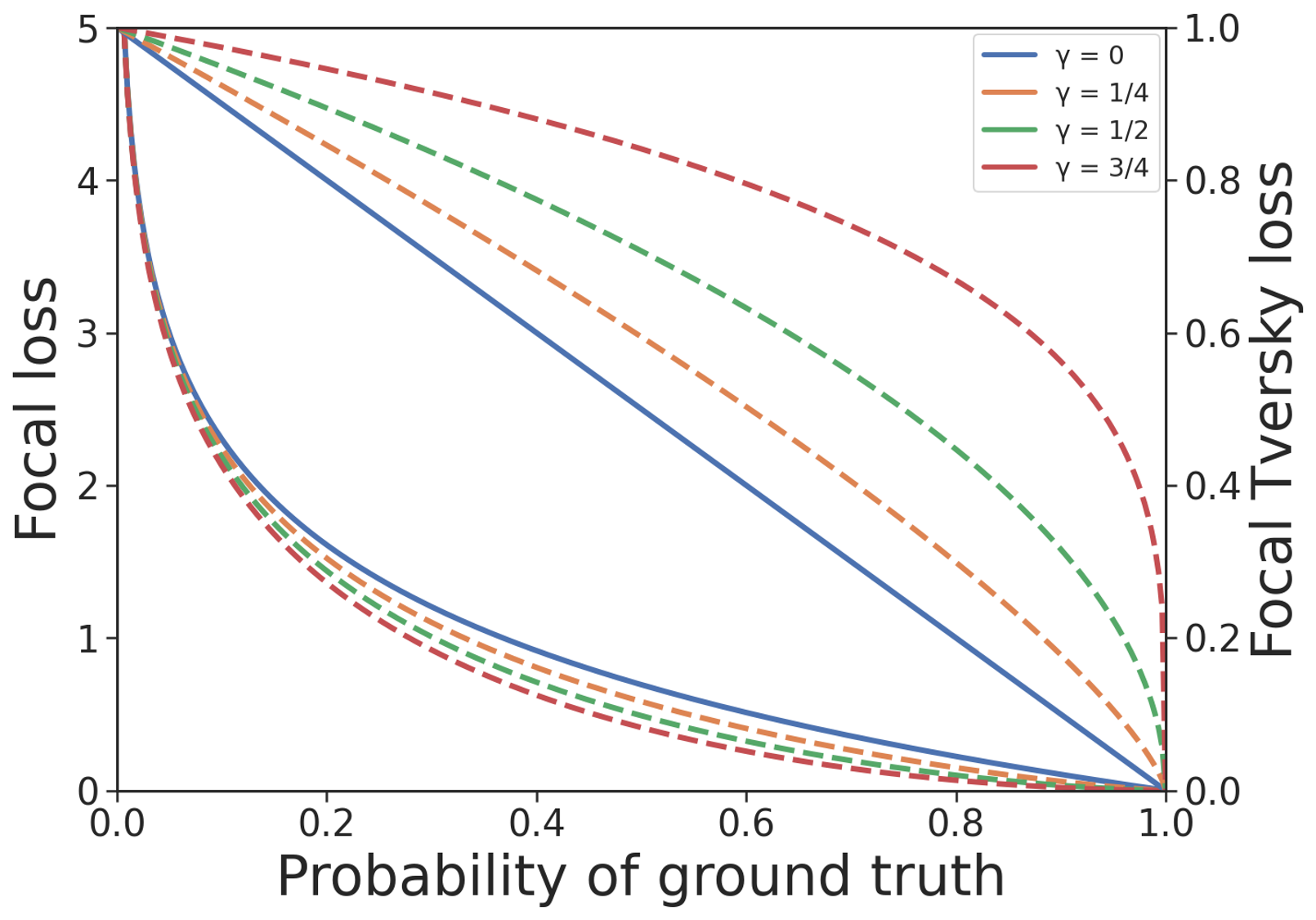}
    \caption{Effect of changing $\gamma$ with the Unified Focal loss. The top and bottom group of curves relate to the Focal Tversky loss and Focal loss respectively. The dashed lines represent different values of $\gamma$ with the modified Focal Tversky loss and modified Focal loss components.}
    \label{fig:figure_2}
\end{figure}

The issue of loss suppression associated with the Focal loss is mitigated by complementary pairing with the Focal Tversky loss, with the asymmetry enabling simultaneous background loss suppression and foreground loss enhancement, analogous to increasing the signal to noise ratio (Fig.~\ref{fig:figure_2}). 

By incorporating ideas from previous loss functions, the Unified Focal loss generalises Dice-based and cross entropy-based loss functions into a single framework. In fact, it can be shown that all Dice and cross entropy based loss functions described so far are special cases of the Unified Focal loss (Fig.~\ref{fig:figure_1}). For example, by setting $\gamma = 0$ and $\delta = 0.5$, the Dice loss and the cross entropy loss are recovered when $\lambda$ is set to $0$ and $1$ respectively. By clarifying the relationship between the loss functions, the Unified Focal loss is much easier to optimise than separately trialling the different loss functions, and it is also more powerful because it is robust to both input and output imbalances. Importantly, given that the Dice loss and cross entropy loss both are efficient operations, and applying the focal parameter adds negligible time complexity, the Unified Focal loss is not expected to significantly increase training time over its component loss functions.

In practice, optimisation of the Unified Focal loss can be further simplified to a single hyperparameter. Given the different effect of the focal parameter on each component loss, the role of $\lambda$ is partially redundant, and therefore we recommend settings $\lambda = 0.5$, which assigns equal weight to each component loss, which is also supported by empirical evidence \citep{taghanaki2019combo}. Furthermore, we recommend setting $\delta = 0.6$, to correct the Dice loss tendency to produce high precision, low recall segmentations with class imbalance. This is less than $\delta = 0.7$ in the Tversky loss, to account for the effect from the cross entropy-based component. This heuristic reduction of the hyperparameter search space to the single $\gamma$ parameter makes the Unified Focal loss both powerful and easy to optimise. We provide further empirical evidence behind these heuristics for the Unified Focal loss in the Supplementary Materials.

\section{Materials and methods}
\label{sec:MatMeth}

\subsection{Dataset descriptions and evaluation metrics}
\label{sec:Datasets}

We select five class imbalanced medical imaging datasets for our experiments: CVC-ClinicDB, DRIVE, BUS2017, KiTS19 and BraTS20. To assess the degree of class imbalance, the percentage of foreground pixels/vowels were calculated per image and averaged over the entire dataset (Table~\ref{tab:Table1}). 

\begin{table}[ht!]
\centering
\caption{Details of datasets and training setup used for our experiments.}
\scalebox{0.75}{
\begin{tabular}{lccccccc}
\hline
Dataset      & Segmentation     & \#Images & Input size     & \#Training & \#Validation & \#Test & \%Foreground \\ \hline
CVC-ClinicDB & Colorectal polyp & 612      & 288 $\times$ 384 $\times$ 3  & 392        & 98           & 122    & 9.3          \\
DRIVE        & Retinal vessel   & 40       & 512 $\times$ 512 $\times$ 3  & 16         & 4            & 20     & 8.7          \\
BUS2017      & Breast tumour    & 163      & 128 $\times$ 128 $\times$ 3  & 104        & 26           & 33     & 4.8          \\
BraTS20      & Enhancing tumour & 342      & 96 $\times$ 96 $\times$ 96   & 219        & 55           & 68     & 0.2          \\
KiTS19       & Kidney / Tumour  & 204      & 80 $\times$ 160 $\times$ 160 & 130        & 33           & 41     & 0.8 / 0.2    \\ \hline
\end{tabular}}
\label{tab:Table1}
\end{table}

\subsubsection{CVC-ClinicDB dataset}

Colonoscopy is the gold-standard screening tool for colorectal cancer, but is associated with significant polyp miss rates, presenting an opportunity to leverage computer-aided systems to support clinicians in reducing the number of polyps missed \citep{kim2017miss}. We use the CVC-ClinicDB dataset, which consists of 612 frames containing polyps with image resolution $288 \times 384$ pixels, generated from 23 video sequences from 13 different patients using standard colonoscopy interventions with white light \citep{bernal2015wm}.

\subsubsection{DRIVE dataset}

Degenerative retinal diseases display characteristic features on fundoscopy that may be used to aid diagnosis. In particular, retinal vessel abnormalities such as changes in tortuosity or neovascularisation provide important clues for staging and treatment planning. We select the DRIVE dataset \citep{staal2004ridge}, which consists of 40 coloured fundus photographs obtained from diabetic retinopathy screening in the Netherlands, captured using 8 bits per colour plane of resolution $768 \times 584$. 33 photographs display no signs of diabetic retinopathy, while 7 photographs show signs of mild diabetic retinopathy.

\subsubsection{BUS2017 dataset}

The most commonly used screening tool for breast cancer assessment is digital mammography. However, dense breast tissue, often seen in younger patients, is poorly visualised on mammography. An important alternative is US imaging, which is an operator-dependent procedure requiring skilled radiologists, but has the advantage of no radiation exposure unlike mammography. BUS2017 dataset B consists of $163$ ultrasound images and associated ground truth segmentations with mean image size is $760\times570$ pixels collected from the UDIAT Diagnostic Centre of the Parc Taul\'i Corporation, Sabadell, Spain. 110 images are benign lesions, consisting of 65 unspecified cysts, 39 fibroadenomas and 6 from other benign types. The other 53 images depict cancerous masses, with the majority invasive ductal carcinomas. 

\subsubsection{BraTS20 dataset}

BraTS20 dataset is currently the largest, publicly available and fully-annotated dataset for medical image segmentation \citep{nazir2021}, and comprises of $494$ multimodal scans of patients with either low-grade glioma or high-grade glioblastoma \citep{menze2014multimodal,bakas2017advancing,bakas2018identifying}. The BraTS20 dataset provides images for the following MRI sequences: T1-weighted (T1), T1-weighted contrast-enhanced using gadolinium contrast agents (T1-CE), T2-weighted (T2) and fluid attenuated inverse recovery (FLAIR) sequence. Images were manually annotated, with regions associated with the tumour labelled as: necrotic and non-enhancing tumour core, peritumoural oedema or gadolinium-enhancing tumour. From the $494$ scans provided, $125$ scans are used for validation with reference segmentation masks withheld from public access, and therefore are excluded. To define a binary segmentation task, we further exclude T1, T2 and FLAIR sequences to focus on gadolinium-enhancing tumour segmentation using the T1-CE sequence \citep{rundo2019MedGA,Han2019}, which not only appears to be the the most difficult class to segment \citep{henry2020top}, but is also the most clinically relevant for radiation therapy \citep{rundoCMPB2017, rundo2018next}. We further exclude another $27$ scans without enhancing tumour regions, leaving $342$ scans, with image resolution $240\times240\times155$ voxels, for use.

\subsubsection{KiTS19 dataset}

Kidney tumour segmentation is a challenging task due to the widespread presence of hypodense tissue, as well as highly heterogeneous appearance of tumours on CT \citep{linguraru2009,rundo2020CBM}. To evaluate our loss functions, we select the KiTS19 dataset \citep{heller2019kits19}, a highly class imbalanced, multi-class classification problem. Briefly, this dataset consists of $300$ arterial phase abdominal CT scans from patients who underwent partial removal of the tumour and surrounding kidney or complete removal of the kidney including the tumour at the University of Minnesota Medical Center, USA. The image size is $512\times512$ pixels in the axial plane, with an average of $216$ slices in coronal plane. Kidney and tumour boundaries were manually delineated by two students, with class labels of either kidney, tumour or background assigned to each voxel resulting in a semantic segmentation task \citep{heller2019kits19}. $210$ scans and their associated segmentations are provided for training, with the segmentation masks for the other $90$ scans withheld from public access for testing. We therefore exclude the 90 scans without segmentation masks, and further exclude another 6 scans (case 15, 23, 37, 68, 125 and 133) due to concern over ground truth quality \citep{heller2019state}, leaving $204$ scans for use. 

\subsubsection{Evaluation metrics}
\label{sec:EvalMetrics}

To assess segmentation accuracy, we use four commonly used metrics \citep{wang2020image}: DSC, Intersection over Union (IoU), recall and precision. DSC is defined in Eq.~(\ref{eq:DSC}), and IoU, recall and precision are similarly defined per pixel/voxel and according to Eqs.~(\ref{eq:iou}), (\ref{eq:recall}) and (\ref{eq:precision}), respectively:

\begin{equation}
\mathrm{IoU}=\frac{\mathrm{TP}}{\mathrm{TP}+\mathrm{FP}+\mathrm{FN}},
\label{eq:iou}
\end{equation}

\begin{equation}
\text {Recall}=\frac{\text{TP}}{\text{TP}+\text{FN}},
\label{eq:recall}
\end{equation}

\begin{equation}
\text {Precision}=\frac{\text{TP}}{\text{TP}+\text{FP}}.
\label{eq:precision}
\end{equation}

\subsection{Implementation details}
\label{sec:Implementation}

All experiments are programmed using Keras with TensorFlow backend and run on NVIDIA P100 GPUs. We made use of the Medical Image Segmentation with Convolutional Neural Networks (MIScnn) open-source Python library \citep{muller2019miscnn}.

Images from the CVC-ClinicDB, DRIVE and BUS2017 datasets are provided in an anonymised tiff, jpeg and png file formats respectively. For both the KiTS19 and BraTS20 dataset, images and ground truth segmentation masks are provided in an anonymised NIfTI file format. For all datasets, except for the DRIVE dataset which is originally partitioned into 20 training images and 20 testing images, we randomly partitioned each dataset into 80\% development and 20\% test set, and further divided the development set into 80\% training set and 20\% validation set. All images were normalised to $[0,1]$ using the z-score. We made use of the `batchgenerators' library to apply on-the-fly data augmentation with probability $0.15$, including: scaling ($0.85-1.25\times$), rotation ($-15\degree$ to $+15\degree$), mirroring (vertical and horizontal axes), elastic deformation ($\alpha \in [0, 900]$ and $\sigma \in [9.0, 13.0]$) and brightness ($0.5-2\times$).

For 2D binary segmentation, we used the CVC-ClinicDB, DRIVE and BUS2017 datasets and perform full-image analysis, with images resized as described in (Table~\ref{tab:Table1}). For 3D binary segmentation, we used the BraTS20 dataset. Here, images were pre-processed, with the skull stripped and images interpolated to the same isotropic resolution of $1$mm$^3$, and we performed patch-wise analysis using random patches of size of $96\times96\times96$ voxels for training with patch-wise overlap of $48\times48\times48$ voxels for inference. For 3D multiclass segmentation, we used the KiTS19 dataset. Hounsfield units (HU) were clipped to $[-79, \ldots, 304]$ HU and voxel spacing resampled to $3.22\times1.62\times1.62$mm$^3$ \citep{muller2019miscnn}. We performed patch-wise analysis using random patches of size of $80\times160\times160$ voxels for training and patch-wise overlap of $40\times80\times80$ voxels for inference.

For the 2D segmentation tasks, we used the original 2D U-Net architecture \citep{ronneberger2015u}, and for the 3D segmentation tasks, we used the 3D U-Net  \citep{cciccek20163d}. Model parameters were initialised using Xavier initialisation \citep{glorot2010understanding}, and we added instance normalisation and a final softmax activation layer \citep{zhou2019normalization}. We trained using the stochastic gradient descent optimiser with a batch size of 2 and initial learning rate of $0.1$. For convergence criteria, we used ReduceLROnPlateau to reduce the learning rate by $0.1$ if the validation loss did not improve after 10 epochs, and the EarlyStopping callback to terminate training if the validation loss did not improve after 20 epochs. Validation loss was evaluated after each epoch, and the model with the lowest validation loss was selected as the final model.

We evaluate the following loss functions: cross entropy loss, Focal loss, Dice loss, Tversky loss, Focal Tversky loss, Combo loss, and symmetric and asymmetric variants of the Unified Focal loss. We used optimal hyperparameters for each loss function as reported in the original studies. Specifically, we set $\alpha=0.25$ and $\gamma=2$ for the Focal loss \citep{Lin_2017_ICCV}, $\alpha=0.3$, $\beta=0.7$ for the Tversky loss \citep{salehi2017tversky}, $\alpha=0.3$, $\beta=0.7$ and $\gamma=4/3$ for the Focal Tversky loss \citep{8759329} and $\alpha = \beta = 0.5$ for the Combo loss. For the Unified Focal loss, we set $\lambda = 0.5$, $\delta = 0.6$, and performed hyperparameter tuning with $\gamma \in [0.1,0.9]$ for the 2D segmentation tasks, and set $\gamma = 0.5$ for the 3D segmentation tasks.

To test for statistical significance, we used the Wilcoxon rank-sum test. A statistically significant difference was defined as $p < 0.05$.

\section{Experimental results}
\label{sec:Results}

In this section, we first describe the results from the 2D binary segmentation using the CVC-ClinicDB, DRIVE and BUS2017 datasets, followed by 3D binary segmentation using the BraTS20 dataset, and conclude with 3D multiclass segmentation with the KiTS19 dataset.

\subsection{2D binary segmentation}

The results for the 2D binary segmentation experiments are shown in Tables~\ref{tab:CVC-ClinicDB}--\ref{tab:BUS2017}.

\begin{table}[ht!]
\caption{Results on the CVC-ClinicDB dataset. Values are in the form mean $\pm$ $95\%$ confidence intervals. Numbers in boldface denote the highest values for each metric. The best values for the Unified Focal losses are reported from the hyperparameter tuning.}
\centering
\scalebox{0.9}{
\begin{tabular}{lcccc}
\hline
Loss function  & DSC                  & IoU                  & Precision            & Recall               \\ \hline
CE             & 0.889±0.025          & 0.820±0.029          & 0.921±0.026          & 0.878±0.026          \\
Focal          & 0.868±0.027          & 0.790±0.031          & 0.844±0.032          & \textbf{0.933±0.017} \\
DSC            & 0.867±0.029          & 0.792±0.034          & 0.895±0.030          & 0.875±0.031          \\
Tversky        & 0.874±0.025          & 0.796±0.030          & 0.864±0.029          & 0.909±0.025          \\
Focal Tversky  & 0.894±0.026          & 0.831±0.030          & 0.896±0.026          & 0.919±0.023          \\
Combo          & 0.895±0.025          & 0.831±0.030          & \textbf{0.927±0.023} & 0.885±0.028          \\
Unified Focal (Sym)  & \textbf{0.909±0.024} & \textbf{0.852±0.028}          & 0.917±0.026          & 0.919±0.020          \\
Unified Focal (Asym) & \textbf{0.909±0.023} & 0.851±0.028 & 0.910±0.026          & 0.932±0.016          \\ \hline
\end{tabular}}
\label{tab:CVC-ClinicDB}
\end{table}

\begin{table}[ht!]
\caption{Results on the DRIVE dataset. Values are in the form mean $\pm$ $95\%$ confidence intervals. Numbers in boldface denote the highest values for each metric. The best values for the Unified Focal losses are reported from the hyperparameter tuning.}
\centering
\scalebox{0.9}{
\begin{tabular}{lcccc}
\hline
Loss function  & DSC                  & IoU                  & Precision            & Recall               \\ \hline
CE             & 0.789±0.008          & 0.652±0.011          & \textbf{0.874±0.017} & 0.742±0.021          \\
Focal          & 0.781±0.008          & 0.642±0.011          & 0.724±0.019          & \textbf{0.853±0.016} \\
DSC            & 0.799±0.007          & 0.666±0.009          & 0.817±0.018          & 0.787±0.021          \\
Tversky        & 0.794±0.007          & 0.658±0.009          & 0.750±0.019          & 0.848±0.018          \\
Focal Tversky  & 0.798±0.007          & 0.664±0.010          & 0.765±0.020          & 0.839±0.019          \\
Combo          & 0.796±0.007          & 0.661±0.010          & 0.836±0.017          & 0.763±0.021          \\
Unified Focal (Sym)  & 0.801±0.006          & 0.669±0.009          & 0.816±0.018          & 0.792±0.021          \\
Unified Focal (Asym) & \textbf{0.803±0.006} & \textbf{0.671±0.008} & 0.793±0.018          & 0.818±0.020          \\ \hline
\end{tabular}}
\label{tab:DRIVE}
\end{table}

\begin{table}[ht!]
\caption{Results on the BUS2017 dataset. Values are in the form mean $\pm$ $95\%$ confidence intervals. Numbers in boldface denote the highest values for each metric. The best values for the Unified Focal losses are reported from the hyperparameter tuning.}
\centering
\scalebox{0.9}{
\begin{tabular}{lcccc}
\hline
Loss function  & DSC                  & IoU                  & Precision            & Recall               \\ \hline
CE             & 0.751±0.086          & 0.653±0.09           & 0.730±0.098          & 0.851±0.075          \\
Focal          & 0.603±0.092          & 0.480±0.090          & 0.659±0.113          & 0.710±0.102          \\
DSC            & 0.767±0.085          & 0.672±0.088          & 0.788±0.094 & 0.808±0.075          \\
Tversky        & 0.808±0.070          & 0.716±0.078          & 0.780±0.081          & 0.904±0.039          \\
Focal Tversky  & 0.799±0.081          & 0.712±0.085          & 0.758±0.090          & 0.912±0.062          \\
Combo          & 0.759±0.087          & 0.665±0.092          & 0.746±0.094          & 0.849±0.073          \\
Unified Focal (Sym)  & 0.814±0.063          & 0.716±0.070          & 0.768±0.076          & \textbf{0.923±0.027} \\
Unified Focal (Asym) & \textbf{0.824±0.063} & \textbf{0.731±0.071} & \textbf{0.797±0.074}          & 0.908±0.037          \\ \hline
\end{tabular}}
\label{tab:BUS2017}
\end{table}

Across all three datasets, the best performance was consistently observed with the asymmetric variant of the Unified Focal loss, achieving a DSC of $0.909 \pm 0.023, 0.803 \pm 0.006$ and $0.824 \pm 0.063$ on the CVC-ClinicDB, DRIVE and BUS2017 datasets respectively. This was followed by the symmetric variant of the Unified Focal loss, which achieved the best IoU score of $0.852 \pm 0.028$ on the CVC-ClinicDB dataset, and comparable DSC scores to the asymmetric variant with a DSC of $0.909 \pm 0.024, 0.801 \pm 0.006$ and $0.814 \pm 0.063$ on the CVC-ClinicDB, DRIVE and BUS2017 datasets. No statistically significant difference in performance was observed between the two variants of the Unified Focal loss on these datasets. Generally, the worst performance was observed with cross entropy-based loss functions, with the Focal loss performing significantly worse than the cross entropy loss on the CVC-ClinicDB ($p=0.04$) and BUS2017 ($p=0.004$) datasets, and significantly worse than the asymmetric variant of the Unified Focal loss across the three datasets (CVC-ClinicDB: $p=\num{2e-6}$, DRIVE: $p=\num{1e-4}$ and BUS2017: $p=\num{5e-5}$). No significant differences were observed between the Dice-based losses.

\begin{figure}[ht!]
    \centering
    \includegraphics[width=1.\textwidth]{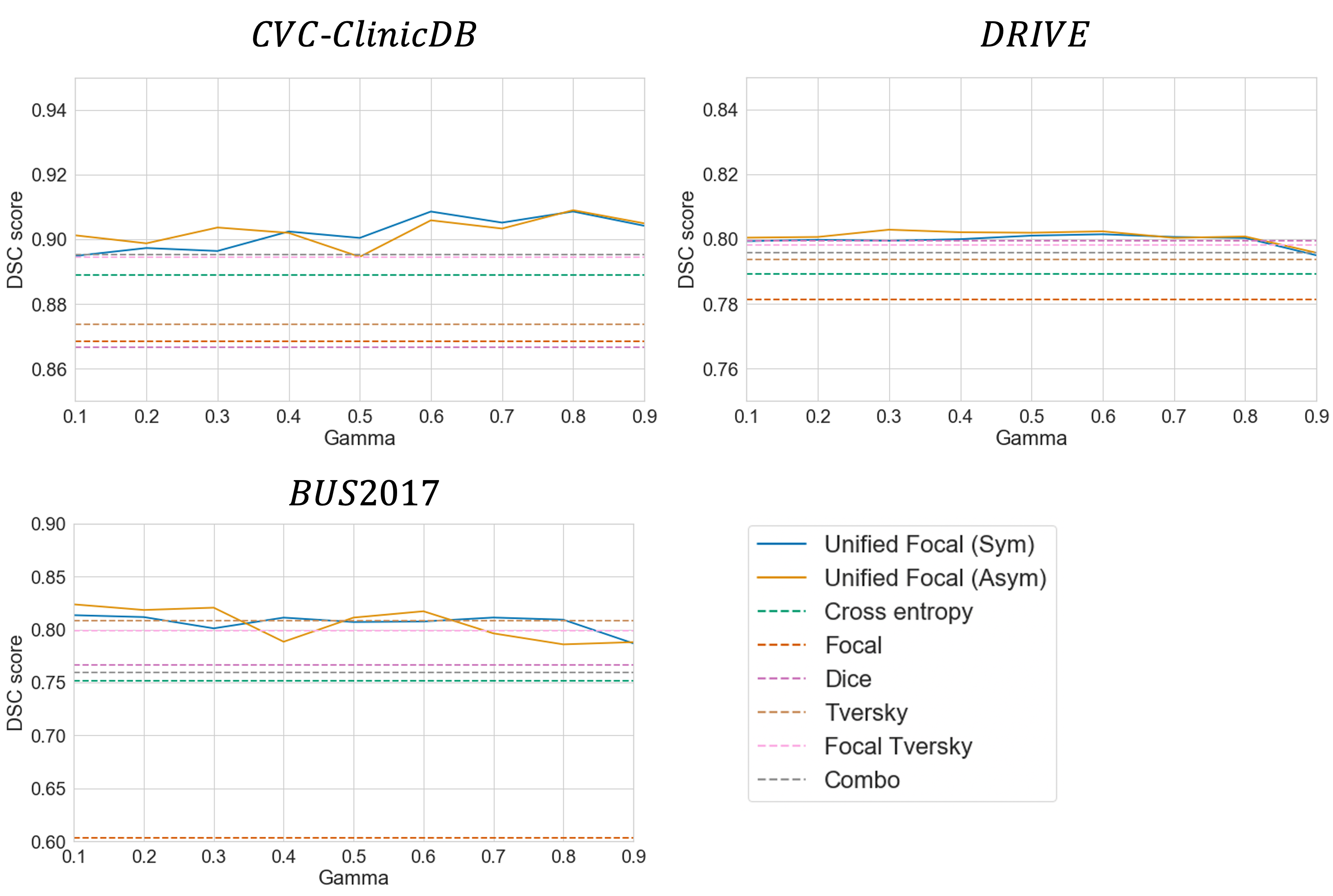}
    \caption{Evaluating the stability of $\gamma$ with the Unified Focal loss on the DSC performance for each dataset. The solid lines represent the symmetric and asymmetric variants of the Unified Focal loss, and for reference, the dashed lines represent the DSC performance of the other loss functions.}
    \label{fig:figure_3}
\end{figure}

To evaluate the performance stability of the $\gamma$ hyperparameter, we display the DSC performance for each value of $\gamma \in [0.1,0.9]$ for the three datasets in Fig.~\ref{fig:figure_3}.

For both the symmetric and asymmetric variants, the Unified Focal loss displays consistently strong performance across the range of $\gamma \in [0.1, 0.9]$. This is most evident with the CVC-ClinicDB dataset, where improved performance over the other loss functions is observed across the entire range of hyperparameter values. The worst performance occurred at high values such as $\gamma = 0.9$, while middle values, such as $\gamma = 0.5$, provided robust performance benefits across datasets.

\begin{figure}[ht!]
    \centering
    \includegraphics[scale=0.4]{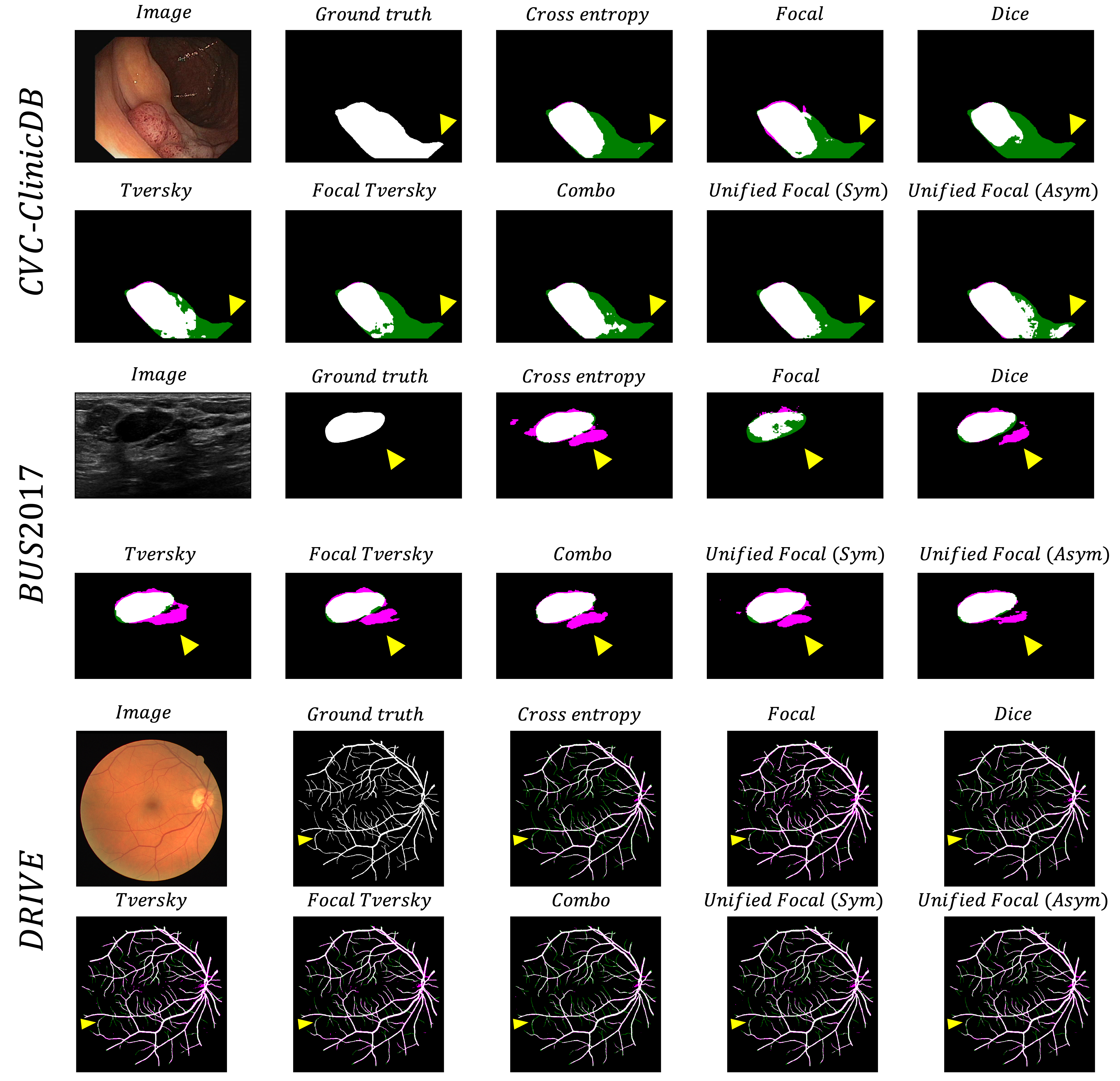}
    \caption{Example segmentations, for each loss function for each of the three datasets. The image and ground truth are provided for reference. The false positive are highlighted in magenta, and the false negatives are highlighted in green. The yellow arrows highlight example areas where segmentation quality differs.}
    \label{fig:figure_4}
\end{figure}

To enable a qualitative comparison, example segmentations are shown in Fig.~\ref{fig:figure_4}.

There is a clear visual difference between the segmentations generated using different loss functions. The segmentations from cross entropy-based loss functions are associated with a greater proportion of false negative predictions compared to the Dice-based loss functions. The highest quality segmentations were produced by the compound loss functions, with the best segmentations produced using the Unified Focal loss. This is particularly clear with the asymmetric variant of the Unified Focal loss in the CVC-ClinicDB example.

\subsection{3D binary segmentation}

The results for the 3D binary segmentation experiments are shown in Tables~\ref{tab:BraTS20}.

\begin{table}[ht!]
\caption{Results on the BraTS20 dataset. Values are in the form mean $\pm$ $95\%$ confidence intervals. Numbers in boldface denote the highest values for each metric. $\gamma = 0.5$ for the Unified Focal losses.}
\centering
\scalebox{0.9}{
\begin{tabular}{lllll}
\hline
Loss function        & DSC                  & IoU                 & Precision            & Recall               \\ \hline
CE                   & 0.716±0.061          & 0.604±0.058         & \textbf{0.826±0.055} & 0.682±0.068          \\
Focal                & 0.738±0.055          & 0.623±0.052         & 0.706±0.055          & 0.805±0.063          \\
DSC                  & 0.620±0.054          & 0.482±0.049         & 0.629±0.064          & 0.683±0.058          \\
Tversky              & 0.580±0.056          & 0.443±0.051         & 0.525±0.061          & 0.740±0.062          \\
Focal Tversky        & 0.747±0.052          & 0.632±0.050          & 0.717±0.056          & \textbf{0.822±0.053} \\
Combo                & 0.686±0.056          & 0.563±0.055         & 0.668±0.063          & 0.757±0.056          \\
Unified Focal (Sym)  & 0.780±0.049          & 0.673±0.049         & 0.803±0.049          & 0.792±0.056          \\
Unified Focal (Asym) & \textbf{0.787±0.049} & \textbf{0.683±0.050} & 0.795±0.048          & 0.800±0.056          \\ \hline
\end{tabular}}
\label{tab:BraTS20}
\end{table}

The best performance was observed with the Unified Focal loss, specifically the asymmetric variant with a DSC of $0.787 \pm 0.049$, IoU of $0.683 \pm 0.050$, precision of $0.795 \pm 0.048$ and recall of $0.800 \pm 0.056$. This was followed by the symmetric variant of the Unified Focal loss, with no significant difference between the two loss functions. In contrast, the asymmetric Unified Focal loss displayed significantly improved performance compared to all the other loss functions (cross entropy loss: $p = 0.02$, Focal loss: $p = 0.03$, Dice loss: $p = \num{6e-10}$, Tversky loss: $p = \num{5e-11}$, Focal Tversky loss: $p = 0.02$ and Combo loss: $p = \num{1e-4}$).

\begin{figure}[ht!]
    \centering
    \includegraphics[scale=0.5]{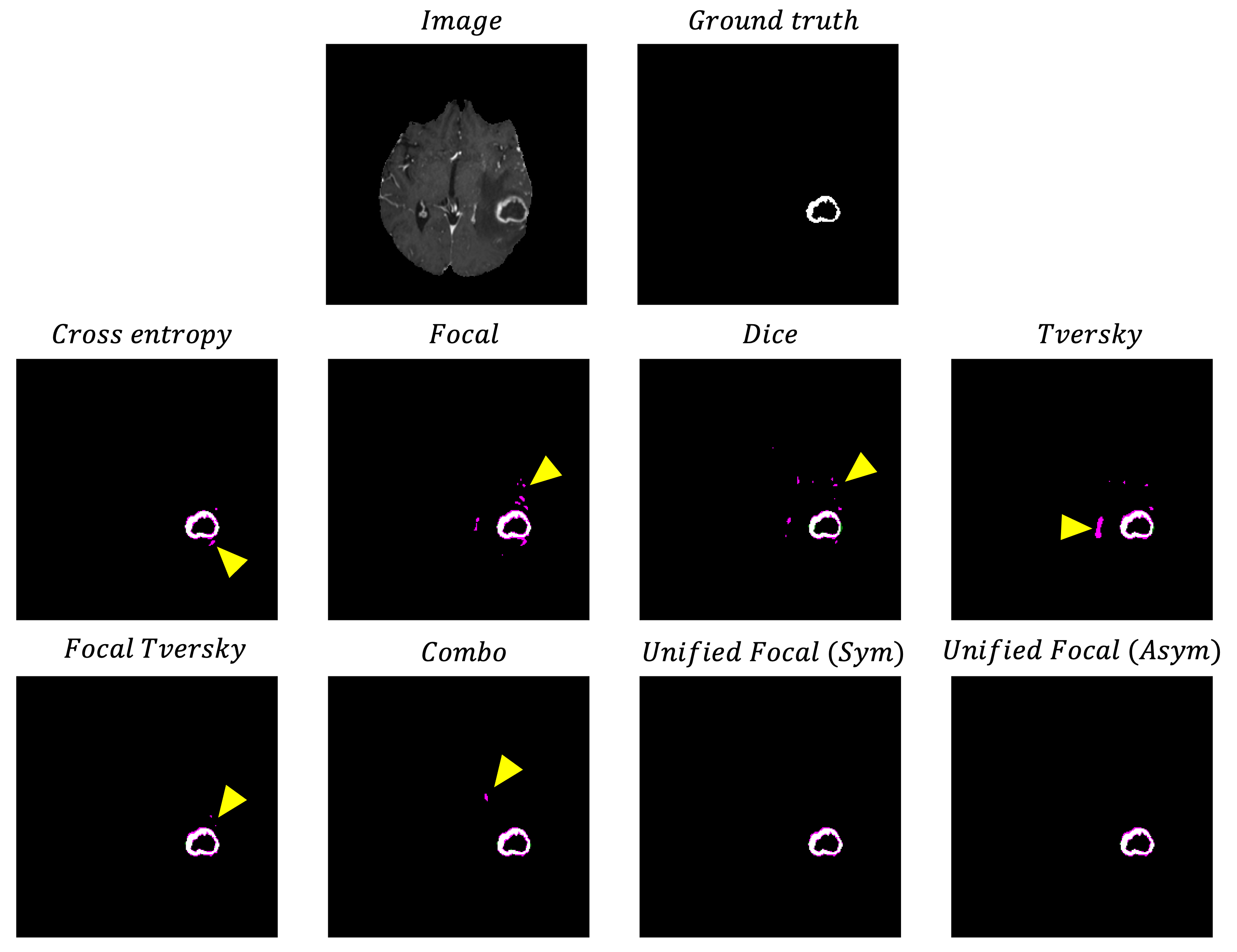}
    \caption{Axial slice from an example segmentation for each loss function for the BraTS20 dataset. The image and ground truth are provided for reference. The false positive are highlighted in magenta, and the false negatives are highlighted in green. The yellow arrows highlight example areas where segmentation quality differs.}
    \label{fig:figure_5}
\end{figure}

Axial slices taken from an example segmentation are shown in Fig.~\ref{fig:figure_5}.

From the results, there is a clear recall bias on this dataset, and this is reflected by the proportion of false positive predictions with each segmentation prediction. The compound loss functions displayed the best recall-precision balance, and this is evident by the significantly reduced false positive predictions visible in the segmentations produced using these loss functions.

\subsection{3D multiclass segmentation}

The results for the 3D multiclass segmentation experiments are shown in Tables~\ref{tab:KiTS19}.

\begin{table}[ht!]
\caption{Results on the KiTS19 dataset. Values are in the form mean $\pm$ $95\%$ confidence intervals. Numbers in boldface denote the highest values for each metric. $\gamma = 0.5$ for the Unified Focal losses.}
\centering
\scalebox{0.5}{
\begin{tabular}{l|llll|llll}
\hline
                     & \multicolumn{4}{c|}{Kidney}                                                                                     & \multicolumn{4}{c}{Tumour}                                                                                     \\ \hline
Loss function        & \multicolumn{1}{c}{DSC} & \multicolumn{1}{c}{IoU} & \multicolumn{1}{c}{Precision} & \multicolumn{1}{c|}{Recall} & \multicolumn{1}{c}{DSC} & \multicolumn{1}{c}{IoU} & \multicolumn{1}{c}{Precision} & \multicolumn{1}{c}{Recall} \\ \hline
CE                   & 0.928±0.016             & 0.869±0.024             & 0.933±0.010                   & 0.928±0.026                 & 0.336±0.107             & 0.262±0.091             & 0.585±0.133                   & 0.303±0.103                \\
Focal                & 0.928±0.011             & 0.868±0.019             & 0.907±0.011                   & \textbf{0.952±0.018}        & 0.349±0.110             & 0.276±0.094             & 0.556±0.136                   & 0.317±0.105                \\
DSC                  & 0.931±0.015             & 0.875±0.024             & 0.936±0.012                   & 0.930±0.024                 & 0.536±0.074             & 0.402±0.069             & 0.594±0.089                   & 0.585±0.077                \\
Tversky              & 0.914±0.017             & 0.846±0.026             & 0.894±0.023                   & 0.940±0.016                 & 0.420±0.087             & 0.308±0.075             & 0.411±0.097                   & 0.616±0.087                \\
Focal Tversky        & 0.926±0.013             & 0.864±0.022             & 0.909±0.017                   & 0.946±0.017                 & 0.520±0.089             & 0.401±0.081             & 0.513±0.095                   & 0.619±0.095                \\
Combo                & 0.935±0.015             & 0.881±0.024             & \textbf{0.954±0.008}          & 0.920±0.025                 & 0.554±0.081             & 0.425±0.074             & 0.616±0.091                   & 0.586±0.088                \\
Unified Focal (Sym)  & \textbf{0.943±0.013}    & \textbf{0.894±0.02}     & 0.949±0.007                   & 0.940±0.021                 & 0.614±0.079             & 0.488±0.077             & \textbf{0.667±0.082}          & 0.657±0.084                \\
Unified Focal (Asym) & \textbf{0.943±0.011}    & \textbf{0.894±0.019}    & 0.942±0.015                   & 0.946±0.014                 & \textbf{0.634±0.079}    & \textbf{0.510±0.078}    & 0.656±0.083                   & \textbf{0.695±0.084}       \\ \hline
\end{tabular}}
\label{tab:KiTS19}
\end{table}

The Unified Focal loss achieves the best performance, with DSC of $0.943 \pm 0.011$ and $0.634 \pm 0.079$ with the asymmetric variant, and DSC of $0.943 \pm 0.013$ and $0.614 \pm 0.079$ with the symmetric variant, for the kidney and kidney tumour segmentation respectively. For kidney segmentation, the asymmetric variant of the Unified Focal loss achieves significantly improved performance compared to the cross entropy loss ($p = 0.03$), Focal loss ($p = 0.004$), Tversky loss ($p = 0.001$), and Focal Tversky loss ($p = 0.03$). The worst performance for kidney segmentation was observed using Dice-based losses, with the Tversky loss followed by the Focal Tversky loss. In contrast, the worst performance for kidney tumour segmentation was observed using cross entropy-based losses, with significantly better DSC performance using the Dice loss compared to the cross entropy loss ($p = 0.01$). For kidney tumour segmentation, the asymmetric variant of the Unified Focal loss achieves significantly better DSC performance compared to the cross entropy loss ($p = \num{6e-5}$), Focal loss ($p = \num{1e-4}$), Dice loss ($p < 0.05$) and Tversky loss ($p = \num{4e-4}$).

Axial slices taken from an example segmentation are shown in Fig.~\ref{fig:figure_5}.

\begin{figure}[ht!]
    \centering
    \includegraphics[scale=0.5]{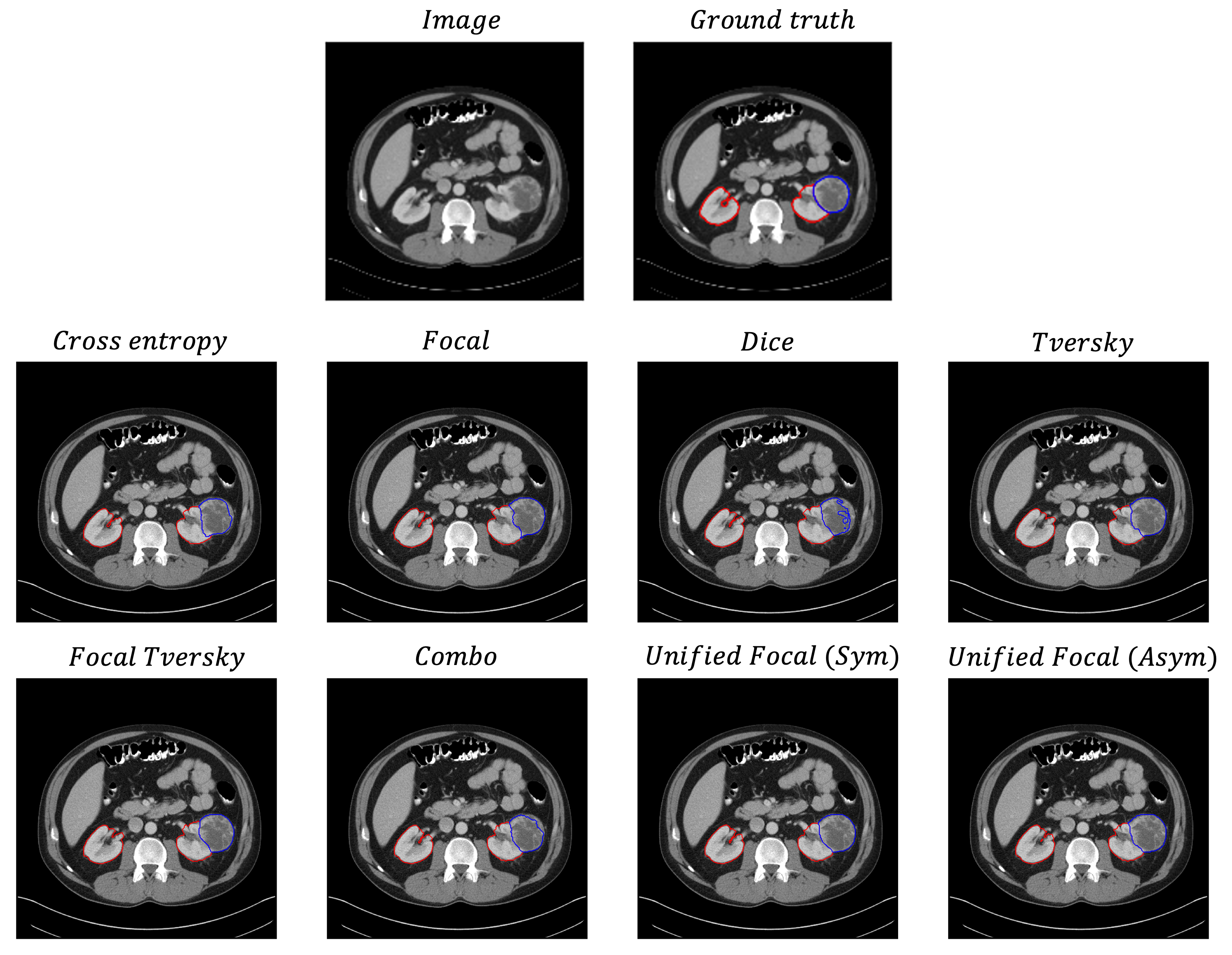}
    \caption{Axial slice from an example segmentation for each loss function for the KiTS19 dataset. The image and ground truth are provided for reference. The red contour corresponds to the kidneys, and the blue contour to the tumour.}
    \label{fig:figure_6}
\end{figure}

While the kidneys are generally well segmented with only subtle differences between the loss functions, the tumour segmentations vary considerably in quality. The low tumour recall scores with the cross entropy-based loss functions are reflected in the segmentations, where the boundary between the tumour and kidney are shifted in favour of kidney prediction. The highest quality segmentation is observed with the Unified Focal loss, with visibly the most accurate contour of the tumour.

\section{Discussion and conclusions}
\label{sec:Discussion}

In this study, we proposed a new hierarchical framework to encompass various Dice and cross entropy-based loss functions, and used this to derive the Unified Focal los, which generalises Dice and cross entropy-based loss functions for handling class imbalance. We compared the Unified Focal loss against six other loss functions on five class imbalanced datasets with varying degrees of class imbalance (CVC-ClinicDB, DRIVE, BUS2017, BraTS20 and KiTS19) involving 2D binary, 3D binary and 3D multiclass segmentation. The Unified Focal loss consistently achieved the highest DSC and IoU scores across the five datasets, with slightly better performance observed using the asymmetric variant over the symmetric variant. We demonstrated that the optimisation of the Unified Focal loss can be simplified to tuning a single $\gamma$ hyperparameter, which we observed is stable and therefore easy to optimise (Fig.~\ref{fig:figure_3}). 

The significant difference in model performance using different loss functions highlights the importance of loss function choice in class imbalanced image segmentation tasks. Most noticeable is the poor performance using distribution-based loss functions with the segmentation of the kidney tumour class on the highly class imbalanced KiTS19 dataset (Table~\ref{tab:KiTS19}). This susceptibility to class imbalance is expected given the greater representation of classes occupying a larger region in cross entropy-based losses. Generally, the Dice-based and compound loss functions performed better with class imbalanced data, but one notable exception was the BraTS20 dataset, where the Dice loss and Tversky loss performed significantly worse than the other loss functions. This likely reflects the unstable gradient issue associated with the Dice loss, resulting in suboptimal convergence and resulting poor performance. Compound loss functions such as the Combo loss and Unified Focal loss performed consistently well across datasets, benefiting from the increased gradient stability with the cross entropy-based component, and the robustness to class imbalance from the Dice-based component. The qualitative assessment correlates with the performance metrics, with the highest quality segmentations observed using the Unified Focal loss (Fig.~\ref{fig:figure_4}--\ref{fig:figure_6}). As expected, no difference in training time was observed between any of the loss functions used in these experiments.

There are several limitations associated with our study. Firstly, we have restricted our framework and comparisons to include only a subset of the most popular variants of the Dice-based and cross entropy-based loss functions. However, it should be noted that the Unified Focal loss also generalises other loss functions that were not included, such as the DiceFocal loss \citep{zhu2019anatomynet} and Asymmetric similarity loss \citep{hashemi2018asymmetric}. One major class of loss functions that were not included were boundary-based loss functions \citep{kervadec2019boundary,zhu2019boundary}, which are another class of loss functions that instead use distance-based metrics to optimise contours rather than distributions or regions used by cross entropy and Dice-based losses, respectively. Secondly, it is not immediately clear how to optimise the $\gamma$ hyperparameter in multiclass segmentation tasks. In our experiments, we treated both the kidney and the kidney tumour as the rare class and assigned $\gamma = 0.5$. Better performance may be observed by assigning different $\gamma$ values to each class, given that for example the kidney class in the KiTS19 dataset is four times more prevalent than the tumour class. However, we still achieved improved performance using the Unified Focal loss over the other loss functions even with this simplification.

We conclude by highlighting several areas for future research. To inform the loss function choice for class imbalanced segmentation, it is important to compare a greater number and variety of loss functions, especially from other loss function classes and with different class imbalanced datasets. We use the original U-Net architecture to simplify but also highlight the importance of loss functions on performance, but it would be useful to assess whether the performance gains generalise to state-of-the-art deep learning methods---such as the nnU-Net~\citep{isensee2020NatMeth}---and whether this is able to complement or even replace alternatives, such as training or sampling-based methods for handling class imbalance.

\section*{Acknowledgements}
This work was partially supported by The Mark Foundation for Cancer Research and Cancer Research UK Cambridge Centre [C9685/A25177], the CRUK National Cancer Imaging Translational Accelerator (NCITA) [C42780/A27066] and the Wellcome Trust Innovator Award, UK [215733/Z/19/Z].
Additional support was also provided by the National Institute of Health Research (NIHR) Cambridge Biomedical Research Centre [BRC-1215-20014] and the Cambridge Mathematics of Information in Healthcare (CMIH) [funded by the EPSRC grant EP/T017961/1].
The views expressed are those of the authors and not necessarily those of the NHS, the NIHR, or the Department of Health and Social Care.

CBS in addition acknowledges support from the Leverhulme Trust project on `Breaking the non-convexity barrier', the Philip Leverhulme Prize, the Royal Society Wolfson Fellowship, the EPSRC grants EP/S026045/1, EP/N014588/1, European Union Horizon 2020 research and innovation programmes under the Marie Skodowska-Curie grant agreement No. 777826 NoMADS and No. 691070 CHiPS, the Cantab Capital Institute for the Mathematics of Information and the Alan Turing Institute.

This work was performed using resources provided by the Cambridge Service for Data Driven Discovery (CSD3) operated by the University of Cambridge Research Computing Service (\url{www.csd3.cam.ac.uk}), provided by Dell EMC and Intel using Tier-2 funding from the Engineering and Physical Sciences Research Council (capital grant EP/P020259/1), and DiRAC funding from the Science and Technology Facilities Council (\url{www.dirac.ac.uk}).


\bibliographystyle{model5-names}
\biboptions{authoryear}

\bibliography{biblio.bib}







\end{document}


\begin{table}[ht!]
\caption{Hyperparameter tuning using the asymmetric variant of the Unified Focal loss on the BUS2017 dataset, confirming the default values chosen for $\delta$ and $\lambda$. Values are reported from five-fold cross validation and in the form mean $\pm$ $95\%$ confidence intervals. Numbers in boldface denote the highest values for each metric.}
\centering
\tiny
\hspace*{-2.2cm}
\scalebox{0.9}{
\begin{tabular}{|cccc|cccc|}
\hline
\multicolumn{4}{|c|}{$\gamma = 0$, $\lambda = 0.5$} & \multicolumn{4}{c|}{$\delta = 0.6$, $\gamma = 0.2$}                                                                               \\ \hline
$\delta$  & DSC         & Precision   & Recall      & $\lambda$ & DSC                                   & Precision                             & Recall                                \\ \hline
0.1       & 0.707±0.055 & \textbf{0.872±0.032} & 0.639±0.063 & 0         & 0.555±0.138                           & 0.764±0.056                           & 0.523±0.163                           \\
0.2       & 0.716±0.034 & 0.852±0.058 & 0.680±0.046 & 0.1       & 0.701±0.061                           & 0.761±0.102                           & 0.743±0.081                           \\
0.3       & 0.716±0.061 & 0.835±0.049 & 0.689±0.093 & 0.2       & 0.704±0.059                           & 0.784±0.067                           & 0.712±0.096                           \\
0.4       & 0.755±0.057 & 0.803±0.061 & 0.766±0.064 & 0.3       & 0.776±0.043                           & 0.775±0.072                           & 0.834±0.030                           \\
0.5       & 0.776±0.058 & 0.821±0.056 & 0.786±0.079 & 0.4       & 0.778±0.020                           & 0.763±0.040                           & \textbf{0.850±0.031} \\
0.6       & \textbf{0.793±0.035} & 0.835±0.061 & 0.814±0.040 & 0.5       & \textbf{0.814±0.027} & \textbf{0.829±0.046} & 0.838±0.038                           \\
0.7       & 0.788±0.056 & 0.774±0.088 & 0.857±0.036 & 0.6       & 0.777±0.034                           & 0.810±0.035                           & 0.806±0.053                           \\
0.8       & 0.755±0.039 & 0.740±0.076 & 0.844±0.040 & 0.7       & 0.771±0.027                           & 0.810±0.046                           & 0.788±0.034                           \\
0.9       & 0.739±0.065 & 0.672±0.092 & \textbf{0.902±0.049} & 0.8       & 0.758±0.034                           & 0.776±0.036                           & 0.813±0.075                           \\
          &             &             &             & 0.9       & 0.735±0.036                           & 0.743±0.112                           & 0.824±0.091                           \\
          &             &             &             & 1         & 0.744±0.042                           & 0.715±0.071                           & 0.843±0.049                           \\ \hline
\end{tabular}}
\label{tab:hyper_BUS}
\end{table}